\documentclass{cimento}

\usepackage{graphicx}

\usepackage{tikz}

\usepackage{isomath} 
\newcommand{\vect}[1]{\vectorsym{#1}}

\usepackage{braket}

\usepackage{siunitx}

\usepackage{bbding}
\definecolor{darkgreen}{cmyk}{1,0,1,0.5}
\newcommand{\cdg}{\color{darkgreen}}
\newcommand{\cdr}{\color{red}}

\newcommand{\CSSM}{Special Research Centre for the Subatomic Structure
  of Matter (CSSM), Department of Physics, University of
  Adelaide, Adelaide, South Australia 5005, Australia}
\newcommand{\UCAS}{School of Physical Sciences, University of Chinese Academy of Sciences (UCAS), Beijing 100049, China}
\newcommand{\SCNT}{Southern Center for Nuclear-Science Theory (SCNT), Institute of Modern Physics, Chinese Academy of Sciences, Huizhou 516000, Guangdong Province, China}
\newcommand{\LanzhouSchool}{School of Physical Science and Technology, Lanzhou University, Lanzhou 730000, China}
\newcommand{\LanzhouHCSR}{Research Centre for Hadron and CSR Physics, Lanzhou University and Institute of Modern Physics of CAS, Lanzhou 730000, China}
\newcommand{\LanzhouTheory}{Lanzhou Center for Theoretical Physics, Key Laboratory of Theoretical Physics of Gansu Province, Key Laboratory of Quantum Theory and Applications of MoE, and Frontiers Science Center for Rare Isotopes, Lanzhou University, Lanzhou 730000, China}
\newcommand{\Julich}{J\"ulich Supercomputing Centre, Forschungszentrum J\"ulich, D-52428 J\"ulich, Germany}

\title{Understanding the nature of baryon resonances}

\author{
Derek B. Leinweber\from{cssm},
Curtis D. Abell\from{cssm},
Liam C. Hockley\from{cssm},\\
\null\ Waseem Kamleh\from{cssm},
Zhan-Wei Liu\from{ls}\from{lh}\from{lt},
Finn M. Stokes\from{cssm}\from{ju},\\
\null\ Anthony W. Thomas\from{cssm},
 \atque
Jia-Jun Wu\from{uc}\from{sc}
}

\instlist{
\inst{cssm}{\CSSM}
\inst{ls}{\LanzhouSchool}
\inst{lh}{\LanzhouHCSR}
\inst{lt}{\LanzhouTheory}
\inst{ju}{\Julich}
\inst{uc}{\UCAS}
\inst{sc}{\SCNT}
}

\shortauthor{D. Leinweber, C. Abell, L. Hockley, W. Kamleh, Z. Liu, F. Stokes, A. Thomas, J. Wu}

\begin{document}

\maketitle

\vspace{-12pt}
\begin{abstract}
This presentation opens with a brief review of lattice QCD
calculations showing the $2s$ radial excitation of the nucleon sits at
approximately 2 GeV, well above the Roper resonance position.  We then
proceed to reconcile this observation with experimental scattering
data.  While the idea of dressing quark-model states in a
coupled-channel analysis to describe scattering data has been around
for decades, it's now possible to bring these descriptions to the
finite-volume of lattice QCD for confrontation with lattice-QCD
calculations.  This combination of lattice QCD and experiment demands
that we reconsider our preconceived notions about the quark-model and
its excitation spectrum.
We close with a discussion of an unanticipated
resolution to the missing baryon resonances problem.
\end{abstract}
\vspace{-36pt}

\section{Introduction}

Perhaps the most surprising and unanticipated discovery in low-lying
baryon spectroscopy is the discovery that the $2s$ radial excitation
of the nucleon lies at approximately 2 GeV
\cite{Mahbub:2010rm,Mahbub:2013ala,Alexandrou:2014mka,%
Leinweber:2015kyz,Roberts:2013ipa,Roberts:2013oea,Kiratidis:2015vpa,%
Kiratidis:2016hda,Khan:2020ahz,Liu:2016uzk,Lang:2016hnn,Stokes:2019zdd},
far from the Roper resonance of the nucleon at 1.44 GeV
\cite{Roper:1964zza}.  So ingrained is the idea that the Roper
resonance is associated with the $2s$ excitation of the simple quark
model, the literature even refers to the $2s$ quark-model excitation
as the Roper state.  We now know this is not the case.

Remarkably the idea that the $\Lambda(1405)$ resonance is not a quark
model state \cite{Hall:2014uca,Hall:2016kou} is widely accepted.  Here
even the two-pole structure associated with attractive coupled-channel
effects in the $\pi \Sigma$ and $\bar K N$ channels is understood
\cite{Molina:2015uqp,Liu:2016wxq,Bulava:2023uma}.

However, this is not necessarily the case for the Roper resonance.
Nevertheless, the Roper resonance should now be understood to be
generated by rescattering in the coupled channels of $\pi N$, $\sigma
N$, and $\pi \Delta$.  Here the $\sigma N$, and $\pi \Delta$ channels
are the resonant contributions from the three-body $\pi \pi N$
channel.

In the following section, we briefly review the lattice QCD
calculations revealing that the radial excitation of the nucleon sits
at 2 GeV. Attempts to find more exotic descriptions of the Roper
resonance in lattice QCD are also reviewed in Sec.~\ref{sec:latt}.

With the radial excitation entrenched at 2 GeV, analysis turns to the
$\pi N$ phase shift data, exploring whether one can describe this data
without a low-lying radial excitation near the Roper resonance.  Here
Hamiltonian Effective Field Theory is used to bring the scattering
data of experiment to the finite volume of lattice QCD to confront the
results of contemporary lattice QCD calculations.  This research is
briefly reviewed in Sec.~\ref{sec:heft}.

These discoveries provide a novel and unanticipated resolution of the
long-standing missing baryon resonances problem and this solution is
presented in the closing section.

\section{Lattice QCD}
\label{sec:latt}

The first hint that there was a problem with the excitation energy of
the $2s$ radial excitation of the nucleon in full $2+1$ flavour
lattice QCD \cite{PACS-CS:2008bkb} was reported in the preprint
of Ref.~\cite{Mahbub:2010rm}.  By using superpositions of Gaussian-shaped smeared
quark sources, this work presents the first attempt to excite the $2s$
radial excitation of the nucleon in full QCD.  

By combining narrow and wide Gaussian sources with opposite signs, one
can create a node in the wave function of the quark distributions
within the nucleon.  The well established generalised eigenvalue
solution \cite{Michael:1985ne,Luscher:1990ck} is employed to calculate
the manner in which the Gaussian-smeared quark sources are superposed
to create the $2s$ radial excitation on the lattice.

Figure 3 of Ref.~\cite{Mahbub:2010rm} illustrates the invariance of
the $2s$ excitation energy obtained.  It was impossible to generate a
low-lying $2s$ excitation.  These results were subsequently confirmed
by the HSC collaboration at heavy quark masses \cite{Edwards:2011jj}
and others \cite{Mahbub:2013ala,Alexandrou:2014mka}. A consensus began
to emerge in 2015 \cite{Leinweber:2015kyz} particularly with the
advent of the Athens Model Independent Analysis Scheme
\cite{Alexandrou:2014mka}.
The wave functions of both the $2s$ and $3s$ excitations were directly
calculated in lattice QCD \cite{Roberts:2013ipa,Roberts:2013oea} and
their profiles compare well with the expectations of a constituent
quark model \cite{Roberts:2013ipa}.

More exotic five-quark descriptions for the Roper resonance were
pursued in Refs.~\cite{Kiratidis:2015vpa,Kiratidis:2016hda}.
However, no new low-lying states were observed.  Similarly the
consideration of hybrid baryons \cite{Khan:2020ahz} did not reveal any
new low-lying states in the Roper channel.

Next generation calculations appeared in 2016 with a high-statistics
calculation coming from the CSSM \cite{Liu:2016uzk} and the very first
insights of the role of two-particle scattering states from Lang {\it
  et al.} \cite{Lang:2016hnn}.  Here the three-quark dominated state
continued to be observed at $\sim 2$ GeV, confirming earlier observations
\cite{Mahbub:2010rm,Mahbub:2013ala,Alexandrou:2014mka,%
Leinweber:2015kyz,Roberts:2013ipa,Roberts:2013oea,Kiratidis:2015vpa,Kiratidis:2016hda},
even when lower-lying scattering states are included in the
correlation matrix.

The role of chiral symmetry in lattice fermion actions was explored
quantitatively in Ref.~\cite{Virgili:2019shg} to see if the explicit
breaking of chiral symmetry in Wilson-clover fermion actions was
responsible for the large $2s$ excitation energy.  A direct 
analysis of lattice correlation functions from Wilson-clover and
overlap fermion actions -- the latter providing a lattice
implementation of chiral symmetry -- revealed no differences in the
spectrum.

Since then, parity-expanded variational analysis (PEVA) techniques
\cite{Stokes:2013fgw} have been developed to explore the
electromagnetic form factors of both the even and odd-parity
excitations of the nucleon \cite{Stokes:2019zdd} and their
electromagnetic transitions to the ground state \cite{Stokes:2019yiz}.
With regard to the $2s$ excitation, the charge radius of the excited
proton is larger than the ground state, in accord with
expectations.  Moreover, the magnetic moments of the excited $2s$
proton and neutron calculated in lattice QCD agree with the
ground-state magnetic moments, again consistent with the expectations
of a $2s$ excitation.

The most recent calculations of the nucleon spectrum are focused on
the lowest-lying scattering states
\cite{Bulava:2023uma,Andersen:2017una,Morningstar:2021ewk,Bulava:2022vpq}.
As described in detail in the next section, these energy levels are
consistent with $\pi N$ scattering data from experiment, providing
confidence in their world-leading computationally-challenging analysis
methods.

In summary, there is overwhelming evidence that the $2s$ radial
excitation of the nucleon has been observed in lattice QCD.  On a
lattice volume of 3 fm on a side, the state sits at 1.9(1) GeV.  
One then naturally asks, is this result even consistent with
experiment?  It is possible to describe $\pi N$ scattering data in the
absence of a low-lying single-particle quark-model state.  If so, then
what is the Roper resonance?

\section{Hamiltonian effective field theory}
\label{sec:heft}

\subsection{Infinite volume}

The idea of dressing quark-model states in a coupled-channel analysis
to describe scattering data has been around for decades
\cite{Thomas:1982kv}.  However, recent advances in understanding the
nature of baryon excitations are flowing from the novel ability to
bring these coupled-channel descriptions to the finite-volume of
lattice QCD for confrontation with lattice-QCD calculations.  This
combination of lattice QCD and experiment demands that we reconsider
our preconceived notions about the quark-model and its excitation
spectrum.

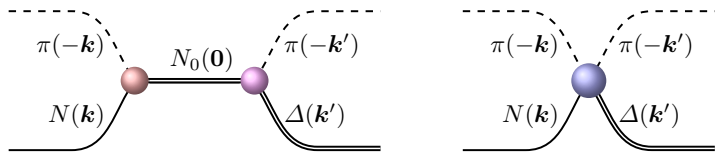
\begin{figure}[t]
\centering
\resizebox{0.7\columnwidth}{!}{%
\begin{tikzpicture}[>=latex,scale=1.0]
  \draw[thick,dashed] (-1.8, 1.0 ) -- (-0.9, 1.0) 
                      to[out=0,  in=125, looseness=1.0] (0.0, 0.0) ;
  \draw[thick       ] (-1.8,-1.0 ) -- (-0.9,-1.0) 
                      to[out=0,  in=-125, looseness=1.0] (0.0, 0.0) ;
  \draw[thick,double] (0.0,0.0) -- (1.5,0.0) ;
  \shade[ball color=red!30!] (0,0) circle (.2) ;
  \draw (-0.3, 0.5) node[left] {$\pi   (-\vect{k})$} ;
  \draw (-0.3,-0.5) node[left] {$N     ( \vect{k})$} ;
\end{tikzpicture}
\hspace{-1.60cm}
\begin{tikzpicture}[>=latex,scale=1.0]
  \draw[thick,double] (-1.5, 0.0) -- ( 0.0, 0.0) ;
  \draw[thick,dashed] (0.0, 0.0) 
                      to[out=55, in=180, looseness=1.0] (0.9, 1.0) 
                      -- (1.8, 1.0) ;
  \draw[thick,double] (0.0, 0.0) 
                      to[out=-55, in=180, looseness=1.0] (0.9,-0.975)
                      -- (1.8,-0.975) ;
  \shade[ball color=magenta!30!] (0,0) coordinate(Hp) circle (.2) ;
  \draw (-0.75,0.0) node[above]{$N_0(\vect{0})$} ;
  \draw ( 0.3, 0.5) node[right]{$\pi(-\vect{k'})$} ;
  \draw ( 0.3,-0.5) node[right]{$\Delta( \vect{k'})$} ;
  \draw ( 1.9, 0.0) node[right]{\null\qquad\null} ;
\end{tikzpicture}
\begin{tikzpicture}[>=latex,scale=1.0]
  \draw[thick,dashed] (-1.8, 1.0) -- (-0.9, 1.0) 
                      to[out=0,  in=125, looseness=1.0] (0.0, 0.0) 
                      to[out=55, in=180, looseness=1.0] (0.9, 1.0) 
                      -- (1.8, 1.0) ;
  \draw[thick       ] (-1.8,-1.0) -- (-0.9,-1.0) 
                      to[out=0,  in=-125, looseness=1.0] (0.0, 0.0) ;
  \draw[thick,double] ( 0.0, 0.0)
                      to[out=-55, in=180, looseness=1.0] (0.9,-0.975) 
                      -- (1.8,-0.975) ;
  \shade[ball color=blue!30!] (0,0) coordinate(Hp) circle (.25) ;
  \draw (-0.3, 0.5) node[left] {$\pi   (-\vect{k} )$} ;
  \draw (-0.3,-0.5) node[left] {$N     ( \vect{k} )$} ;
  \draw ( 0.3, 0.5) node[right]{$\pi   (-\vect{k'})$} ;
  \draw ( 0.3,-0.5) node[right]{$\Delta( \vect{k'})$} ;
\end{tikzpicture}
}
\caption{Examples of one-to-two (left) and two-to-two (right) vertices
  contributing to $\pi N$ scattering in the rest frame.  Here $N_0$
  denotes the single-particle $2s$ radial excitation of the nucleon. 
\vspace{-18pt}
}
\label{fig:vertices}
\end{figure}

Herein, the infinite volume world of experiment and the finite-volume
world of lattice-QCD are bridged by Hamiltonian effective field theory
(HEFT), a nonperturbative extension of effective field theory
incorporating the L\"uscher formalism \cite{Wu:2014vma,Hall:2013qba}.  HEFT
calculations typically commence in infinite volume with an aim to
describe experimental scattering data
\cite{Liu:2016uzk,Liu:2015ktc,Wu:2017qve,Abell:2021awi,Abell:2023qgj,Abell:2023nex}.
Single and non-interacting two-particle states mix via one-to-two and
two-to-two vertices as depicted in Fig.~\ref{fig:vertices}.

The functional form of the one-to-two vertices are governed by
heavy-baryon chiral perturbation theory
\cite{Hall:2013qba,Abell:2021awi} and the two-to-two vertices are
described by separable potentials which facilitate a closed form
solution of the standard coupled-channel equations
\cite{Abell:2023qgj}.  Here the potentials become phenomenological,
adopting a momentum dependence as demanded by the scattering data
\cite{Liu:2015ktc}.  Remarkably, the L\"uscher formalism embedded
within HEFT ensures the low-lying finite-volume spectrum is
independent of the phenomenology used to describe the data, provided
the data is described accurately \cite{Abell:2021awi}.

Of course HEFT can be used in the more traditional manner where
lattice QCD results constrain the Hamiltonian and infinite-volume
scattering observables are predicted \cite{Li:2019qvh,Li:2021mob}.  In
the baryon sector, there is a lack of precision results showing the
subtle shifts of the non-interacting spectra due to the finite volume
of the lattice.  It is better to bring the precision
experimental scattering data to the finite volume of lattice QCD.

As one moves up in the spectrum, more coupled channels come into
effect.  Reference to the partial decay widths of the resonances under
investigation can inform the essential channels to be included in the
calculation.  For example, the $\pi N$, $\sigma N$ and $\pi \Delta$
channels are all observed in the partial decay widths of the Roper
resonance making their inclusion essential to understanding the
structure of the Roper resonance.  In the absence of experimental
analyses of the $\pi N$ to $\pi \Delta$ and $\sigma N$ scattering
amplitudes, there is insufficient information to uniquely constrain
the Hamiltonian from experimental data alone.

As a result, fits to the experimental data provide a variety of
Hamiltonians, all describing the experimental $\pi N$ phase shift and
inelasticity but describing the composition of the spectrum in a
different manner.  It is here, that the finite-volume predictions of
the various coupled-channel fits can be confronted with lattice QCD
results.

\subsection{Finite volume}

In going from infinite volume to finite volume, the continuum of
momentum states becomes quantised to the momenta available on a finite
periodic volume.  Scattering potentials constrained by experiment pick
up finite-volume factors and form the elements of a matrix
Hamiltonian.  One then solves the Schr\"odinger equation
\begin{equation}
  \langle\, i \,|\, H \,|\, j \,\rangle \: \langle\, j \,|\, E_\alpha \,\rangle 
= E_\alpha \, \langle\, i \,|\, E_\alpha \,\rangle \, ,
\end{equation}
where $|\, i \,\rangle$ and $|\, j \,\rangle$ are the non-interacting
basis states taking the discrete momenta available on the periodic
volume ({\it e.g.}  $|\, \pi(\vect k)\, N(-\vect k) \,\rangle\, $),
$E_\alpha$ is the energy eigenvalue for the finite-volume energy
eigenstate labelled by $\alpha$, and $\langle\, i \,|\, E_\alpha
\,\rangle$ is the eigenvector of the Hamiltonian matrix $\langle\, i
\,|\, H \,|\, j \,\rangle$ describing the composition of the
finite-volume energy eigenstate $\alpha$ in terms of the
non-interacting basis states $|\, i \,\rangle$.

The ability of HEFT to describe the pion mass dependence of the
finite-volume energy eigenstates and provide a description of their
composition is key to the advances being made.  Of particular
importance is the contribution of the single-particle basis state to
the energy eigenstates.  As analyses of the nucleon spectrum in the
excitation regime of the Roper are performed with local three-quark
interpolating fields, the overlap of the single-particle basis state
$|\, N_0 \,\rangle$ with the energy eigenstate $|\, \alpha \,\rangle$
provides information of the energy eigenstates most likely to be
excited in lattice QCD calculations.

More quantitatively, B\"ar {\it et al.} \cite{Bar:2016jof} provided a
$\chi$PT estimate of the coupling between a smeared nucleon
interpolating field and a non-interacting pion-nucleon basis state
\begin{equation}
  \frac{3}{16}\, \frac{1}{(f_{\pi}\,L)^2\, E_{\pi}\,L}\left( \frac{E_{N} - m_{N}}{E_{N}} \right) \approx 10^{-3}\,,
\end{equation}
where $E_{\pi}$ and $E_{N}$ are on-shell pion and nucleon energies.
The numerical estimate is based on a 3 fm lattice and the lowest
nontrivial momentum contribution where the coupling is largest.  Here
the $1/L^{3}$ dependence of the coupling is manifest as the
non-interacting two-particle momentum state is spread uniformly
throughout the lattice volume.

Noting the small magnitude of the overlap between the local
interpolating field and the two-particle basis states, one concludes
that the state excited by the local interpolating field is the only
local state in the Hamiltonian basis, the single-particle basis state,
$\ket{N_0}$.  Thus, we associate the three-quark nucleon interpolating
field $\bar{\chi}$ acting on the QCD vacuum, $\ket{\Omega}$, with the
bare $2s$-nucleon basis state of HEFT, via $\bar{\chi}(0) \ket{\Omega}
= \ket{N_0}$.  The most likely energy eigenstate to be observed in
lattice QCD is the energy eigenstate having the largest overlap with
$\ket{N_0}$ as indicated by a survey of $\langle\, N_0 \,|\, E_\alpha
\,\rangle$ over $\alpha$.

Remarkably, there is sufficient information in the eigenvectors of
HEFT to actually simulate what lattice QCD correlation functions of
three-quark operators look like
\begin{equation}
  G_{N_0}(t) = \sum_{i} \left| \braket{ N_0 | E_i} \right|^2 e^{-E_i t}\,.
\end{equation}
This was examined in Ref.~\cite{Abell:2023qgj} and the favourable
comparison with lattice QCD results is noteworthy.

\subsection{Confrontation with Lattice QCD}

Noting that experimental $\pi N$ scattering data alone is not enough to
uniquely constrain the Hamiltonian, we bring the coupled-channel
predictions to the finite volume of lattice QCD to confront the
predictions with lattice QCD determinations of the mass spectrum.

Of critical importance is where the states excited by local
three-quark interpolating fields lie in the spectrum.  HEFT not only
predicts the positions of the energy eigenstates but also their
composition.  A careful comparison of the composition in HEFT with the
interpolating fields used to excite the states in lattice QCD is very
powerful in discriminating between various descriptions of the
experimental scattering data.  To be clear, all coupled-channel
analyses considered herein describe the experimental $\pi N$
scattering data well and generate poles in the complex plane in
agreement with the particle data group
\cite{ParticleDataGroup:2022pth}.

To differentiate between the many possible descriptions of
experimental data alone, we label two fits of interest by the bare
mass, $m_0$, of the single-particle state associated with the $2s$
radial excitation of the quark model, namely $m_0 = 1.7$ GeV and $m_0
= 2.0$ GeV.  The former value leads to the familiar scenario where the
quark-model state is dressed by meson-baryon states to lie in the
regime of the Roper resonance.

\begin{figure}[t]
\centering
\includegraphics[width=0.49\columnwidth]{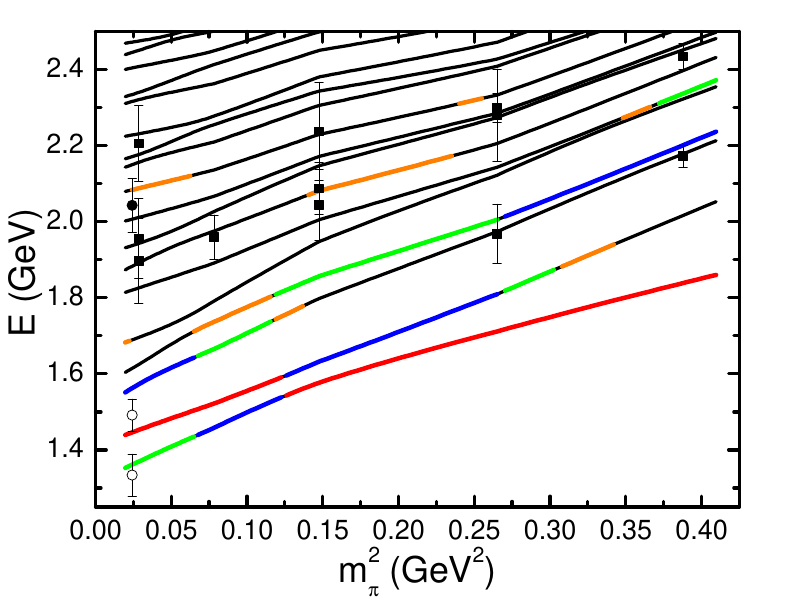}
\includegraphics[width=0.49\columnwidth]{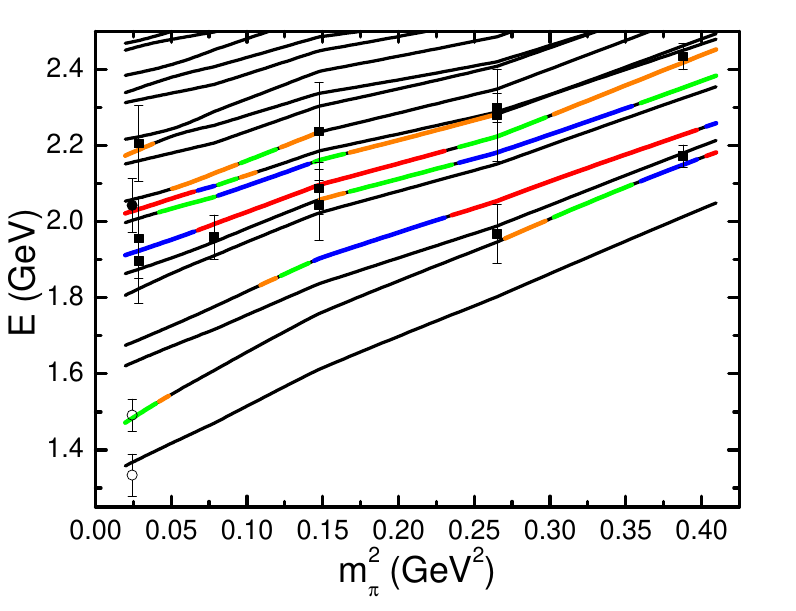}
\vspace{-6pt}
\caption{The finite volume spectra predicted by HEFT \cite{Wu:2017qve}
  (solid lines) for a lattice QCD volume of (3 fm)${}^3$ are
  confronted with results from lattice QCD (bullets and squares).  The
  left plot is the familiar scenario with $m_0 = 1.7$ GeV and the
  right-hand plot has $m_0 = 2.0$ GeV.  Line colours illustrate the
  energy eigenstates with the largest quark-model single-particle
  components in the order red, blue, green, and orange.  These are the
  most likely states to be excited with local three-quark interpolating
  fields in lattice QCD.  Bullets indicate the lattice results from
  Lang {\it et al.} \cite{Lang:2016hnn} and squares are from the CSSM
  \cite{Liu:2016uzk}.  Open symbols denote energy eigenstates
  dominated by two-particle momentum projected interpolating fields
  and full symbols denote energy eigenstates dominated by local
  three-quark interpolating fields.
\vspace{-12pt}
}
\label{fig:comp}
\end{figure}

The excellent description of the $\pi N$ scattering data for both of
these scenarios is presented in Fig.~1 of Ref.~\cite{Wu:2017qve} where
there is very little to differentiate between the two fits.  The
differences become apparent in the finite volume of the lattice. 
These HEFT predictions of the eigenstate energies are presented in
Fig.~\ref{fig:comp}.

The left-hand plot of Fig.~\ref{fig:comp} with $m_0 = 1.7$ GeV
illustrates the prediction of a low-lying state dominated by a
single-particle quark-model like state in the Roper regime, as
indicated by the red curve from HEFT.  The problem with this fit is
that this single-particle dominated state in the regime of the Roper
resonance is not seen in lattice QCD.  While there is a lattice QCD
point on the red curve at $\sim 1.5$ GeV, this state was
excited by a momentum projected $\pi N$ interpolating field
\cite{Lang:2016hnn}. It is not dominated by a three-quark operator.

The right-hand plot of Fig.~\ref{fig:comp} with $m_0 = 2.0$ GeV is in
accord with lattice QCD results.  Here the states predominantly
excited by three-quark interpolating fields are associated with HEFT
eigenstates with large single-particle components indicated by colour
on the eigenstate energy lines.  

Moreover, the composition of the scattering states predicted in HEFT
with $m_0 = 2.0$ GeV match the interpolating fields used to excite the
states in lattice QCD.  In both calculations, the lowest excitation at
$\sim 1.35$ GeV is dominated by $\sigma N$ at zero back-to-back
momentum and the next excitation at $\sim 1.5$ GeV is dominated by
$\pi N$ with the lowest nontrivial back-to-back momentum.  
Table \ref{tab:score} presents a scorecard to evaluate the quality of
these two HEFT descriptions.

\begin{table}[b]
\caption{
  Scorecard for the agreement between lattice QCD calculations and two
  HEFT descriptions of experimental $\pi N$ scattering phase shifts
  and inelasticities.  The $m_0 = 2.0$ GeV description characterises
  the Roper resonance as arising from coupled channel rescattering.
\vspace{-6pt}
}
\label{tab:score}
\begin{tabular}{p{0.6\linewidth}cc}
\hline
\noalign{\smallskip} 
Criteria &$m_0 = 1.7$ GeV  &$m_0 = 2.0$ GeV \\
\noalign{\smallskip}
\hline
\noalign{\smallskip}
Describes experimental data well and produces poles in accord with PDG.  
&{\cdg \CheckmarkBold} &{\cdg \CheckmarkBold} \\
1st lattice scattering state created via $\sigma N$ interpolator has dominant $\langle\, \sigma N \,|\, E_1\,\rangle$ in HEFT.
&{\cdg \CheckmarkBold} &{\cdg \CheckmarkBold} \\
2nd lattice scattering state created via $\pi N$ interpolator has dominant $\langle\, \pi N \,|\, E_2\,\rangle$ in HEFT.
&{\cdr \XSolidBrush}   &{\cdg \CheckmarkBold} \\
Lattice-QCD states excited with 3-quark interpolators are associated with HEFT states with large $\langle\, {N_0} \,|\, E_\alpha\,\rangle$.
&{\cdr \XSolidBrush}   &{\cdg \CheckmarkBold} \\
HEFT predicts three-quark states existing in lattice QCD.
&{\cdr \XSolidBrush}   &{\cdg \CheckmarkBold} \\
\noalign{\smallskip}
\hline
\end{tabular}
\vspace{-12pt}
\end{table}

\begin{table}[t]
\caption{Quark-model predictions \cite{Capstick:1992th} for
  the energies of the radial excitations of the nucleon and the
  $\Delta$ in units of GeV are compared with contemporary
  lattice QCD calculations
  \cite{Roberts:2013oea,Liu:2016uzk,Hockley:2023yzn}.
\vspace{-12pt}
}
\label{tab:qm}
\begin{tabular}{cS[table-format=1.2]S[table-format=1.2]cS[table-format=1.2]S[table-format=1.2]}
\hline
\noalign{\smallskip}
 {State} &{Quark Model}  &{Lattice QCD} &{State} &{Quark Model}  &{Lattice QCD} \\
\noalign{\smallskip}
\hline
\noalign{\smallskip}
$N 1/2^+\ 2s$ &1.54  &1.90(6)  &$\Delta 3/2^+\ 1s$ &1.23  &1.26(2)   \\
$N 1/2^+\ 3s$ &1.77  &2.60(7)  &$\Delta 3/2^+\ 2s$ &1.80  &2.14(5)   \\
$N 1/2^+\ 4s$ &1.88  &3.60(5)  &$\Delta 3/2^+\ 3s$ &1.92  &3.10(17)  \\
$N 1/2^+\ 5s$ &1.98  &         &$\Delta 3/2^+\ 4s$ &1.99  &          \\
\noalign{\smallskip}
\hline
\end{tabular}
\vspace{-12pt}
\end{table}

In the valid HEFT description where $m_0 = 2.0$ GeV, the $\pi \Delta$
channel takes on an enhanced role \cite{Wu:2017qve} in describing the
experimental $\pi N$ phase shift and inelasticity.  In this case the
coupling is large and combines with $\pi N$ and $\sigma N$ channels to
generate a pole in the complex plane.  The Roper resonance arises from
coupled channel rescattering in the $\pi N$, $\sigma N$ and $\pi
\Delta$ channels, the latter two being the resonant contributions of
the three-body $\pi \pi N$ contribution.  This is the nature of the
Roper resonance.

\section{Resolution of the missing baryon resonances problem}
\label{sec:missing}

We close by discussing the impact of the $2s$ radial
excitation of the nucleon being observed at 1.9 GeV in lattice QCD
calculations near the physical point
\cite{Roberts:2013oea,Liu:2016uzk,Lang:2016hnn} as opposed to 1.44 GeV
as anticipated historically \cite{Isgur:1978wd,Capstick:1992th}.

As the single-particle quark model state is mixed with nearby
two-particle states to form resonances, we anticipate the $2s$
radial excitation is largely associated with the $N1/2^+(1880)$ resonance
observed in photoproduction and to a smaller extent the $N1/2^+(1710)$ as
it is only 170 MeV away.  This is supported by the right-hand plot of
Fig.~\ref{fig:comp} where the large single-particle basis state
contributions are illustrated in colour.

In this light, it seems erroneous to tune the parameters of the quark
model to place the radial excitation of the nucleon at 1440 MeV.  The
proliferation of radial excitations below 2 GeV is a direct
consequence of this misidentification.
Table \ref{tab:qm} reports the predictions of the quark model
\cite{Capstick:1992th} tuned to place the $2s$ excitation at 1540 MeV,
a little higher than the Roper resonance in anticipation of some
meson-baryon dressings.  These predictions for the nucleon and
$\Delta$ contrast the reality of lattice QCD calculations.

Had the quark model been tuned to place the $2s$ excitation at 1.9
GeV, the next excitations would be well above 2 GeV.  In this regard,
the missing resonance problem may be regarded as further evidence that
the $2s$ radial excitation of the nucleon is not associated with the
Roper resonance.
\vspace{-3pt}

\acknowledgments
\vspace{-3pt}
This research was undertaken with the assistance of resources from the National Computational
Infrastructure (NCI), and by the Phoenix HPC service at the
University of Adelaide.
This research was supported by the Australian Research Council through ARC Discovery Project Grants Nos. DP190102215 and DP210103706 (D.B.L.).
J.-J. Wu was supported by the National Natural Science Foundation of China under Grant Nos. 12175239 and 12221005, and by the National Key R\&D Program of China under Contract No. 2020YFA0406400.
Z.-W. Liu was supported by the National Natural Science Foundation of China under Grant Nos. 12175091, 11965016, 12047501, and 12247101, and the 111 Project under Grant No. B20063.
\vspace{-6pt}

\bibliographystyle{varenna.bst}
\bibliography{refs}

\end{document}